\documentclass[11pt,twoside]{article}

\usepackage{asp2004}
\usepackage{epsf}
\usepackage{psfig}
\usepackage{lscape}
\def\kms{km~s$^{-1}$}
\usepackage{graphicx}
\begin{document}
\title{The spectrum of the B[e] star BAL224.}   
\author{C. Martayan, M. Floquet, A.-M. Hubert}   
\affil{GEPI/UMR8111-Observatoire de Meudon, 5 place Jules Janssen 92195 Meudon
cedex FRANCE}    
\author{J. Zorec}   
\affil{Institut d'Astrophysique de Paris, 98bis, bd Arago 75014 Paris  FRANCE}    

\begin{abstract} 
We present optical spectroscopy of the emission line star BAL 224 (V=17.3,
B-V=0.46). This star also named KWBBE 485, [MA93]906 is located at the
periphery of the young SMC cluster NGC 330; it is known as
a photometric variable with a possible period around 1 day (Balona 1992).
Furthermore it was reported as the optical counterpart of the prominent
mid-infrared source (MIR1) by Kucinskas et al. (2000), indicating the presence
of a dust shell. The star was included in a sample of B-type stars observed
using the ESO VLT-FLAMES facilities.  The presence of emission lines such as Fe II,[Fe II], [S II] make this
object like a B[e] star. The  H$\alpha$, H$\gamma$ and H$\delta$ lines show an
asymmetrical double-peaked emission profile suggesting the presence of an
accretion disk. Moreover the MACHO and OGLE light curves were analyzed; in
addition to a long-term variability ($\simeq$ 2300d), a short period very close to 1
day has been detected using different methods, confirming the variability
previously reported by Balona (1992). Finally the nature of this object is
reconsidered. 
\end{abstract}


\begin{figure*}[!ht]
\centering
\vskip 0.5cm
\includegraphics[width=6cm, height=13cm,angle=-90]{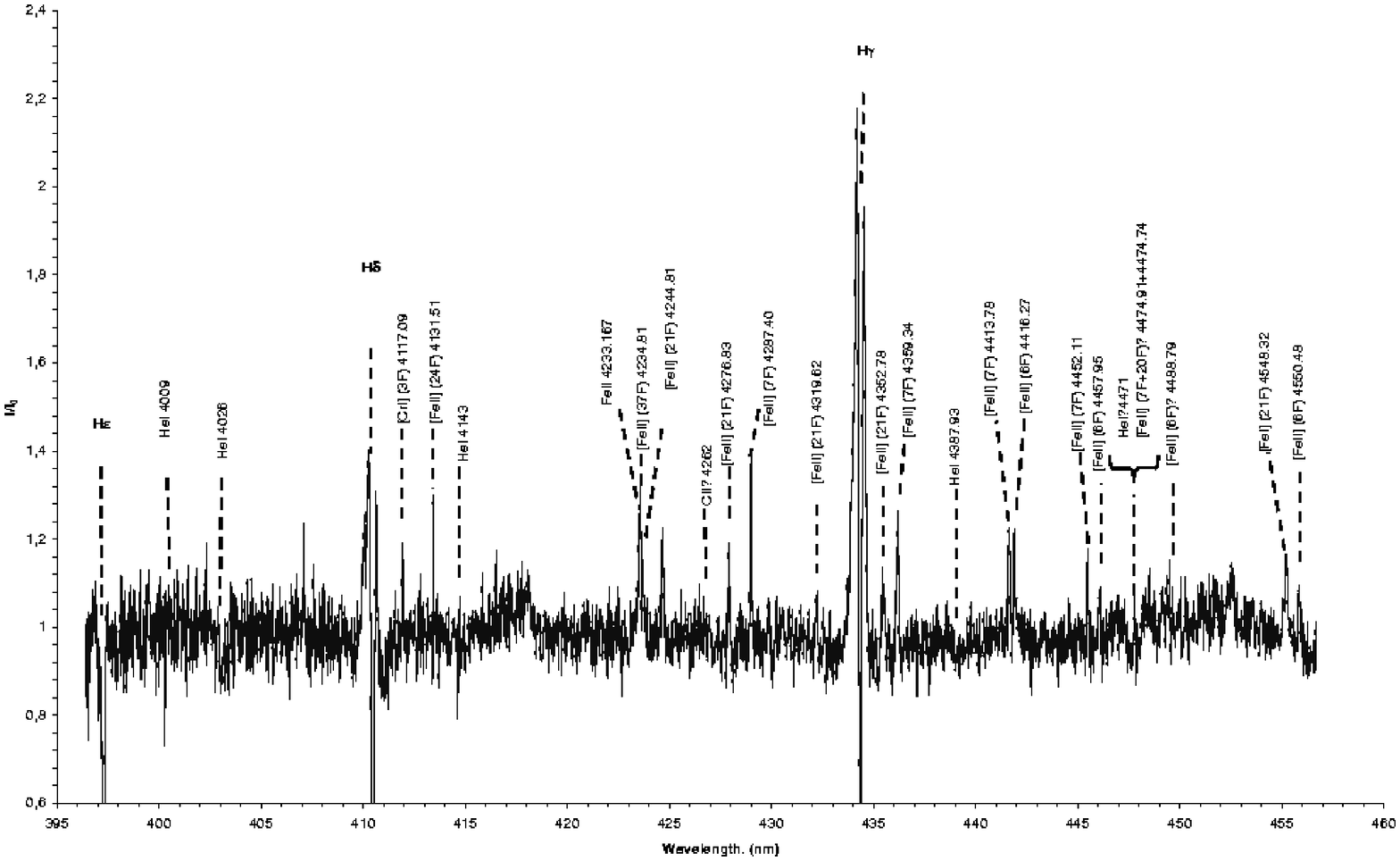}
\vskip 1.5cm
\includegraphics[width=6cm, height=13cm,angle=-90]{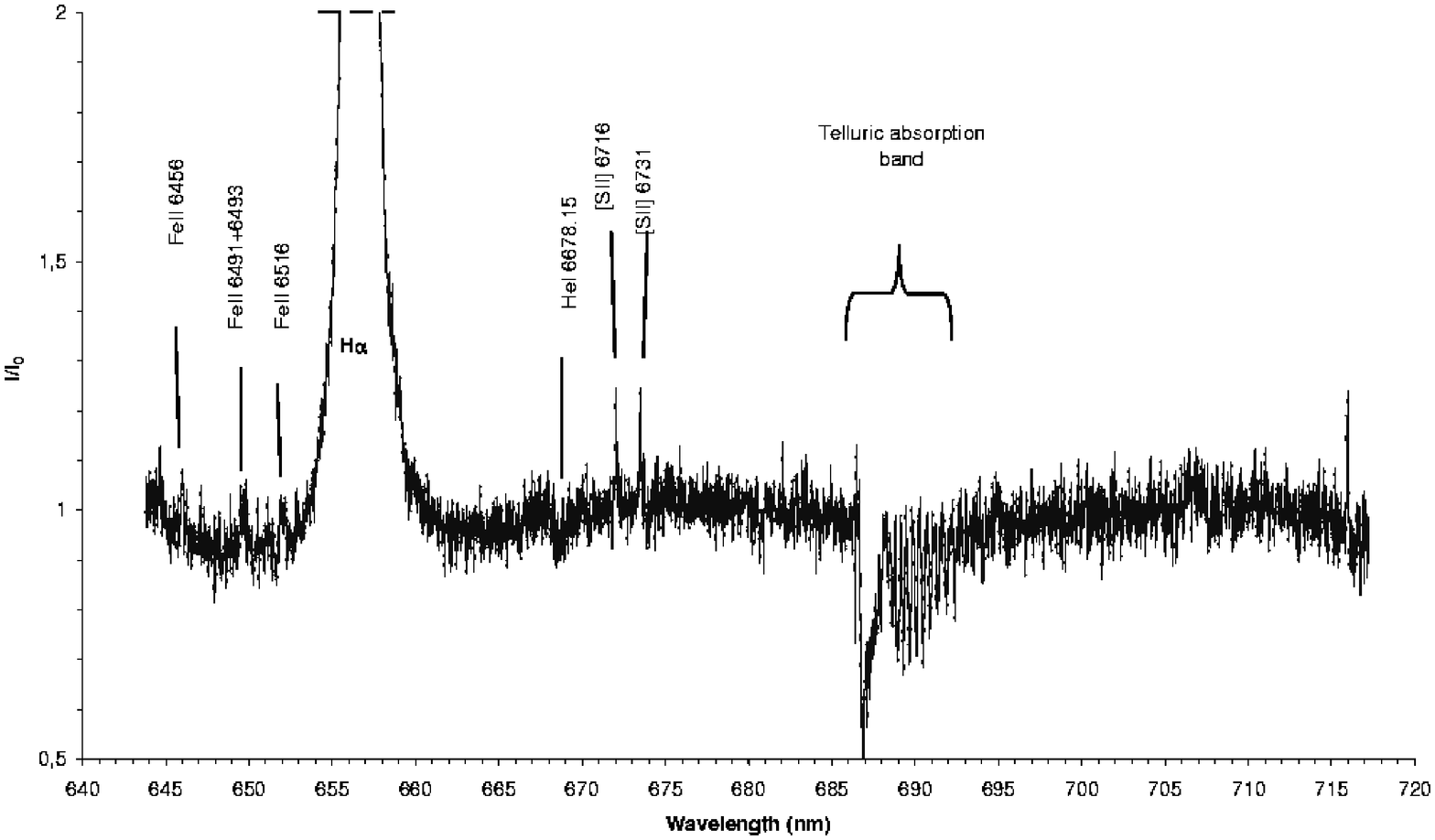}
\caption{Top: Spectrum of the BAL224 in the LR02 VLT-GIRAFFE setting (R=6400). 
Bottom: Spectrum of the BAL224 in the LR06 VLT-GIRAFFE setting (R=8600).}
\label{specL26}
\end{figure*}

\section{Spectroscopic observations}
\begin{figure}[ht]
\centering
\vskip 0.5cm
\includegraphics[width=5cm, height=10cm,angle=-90]{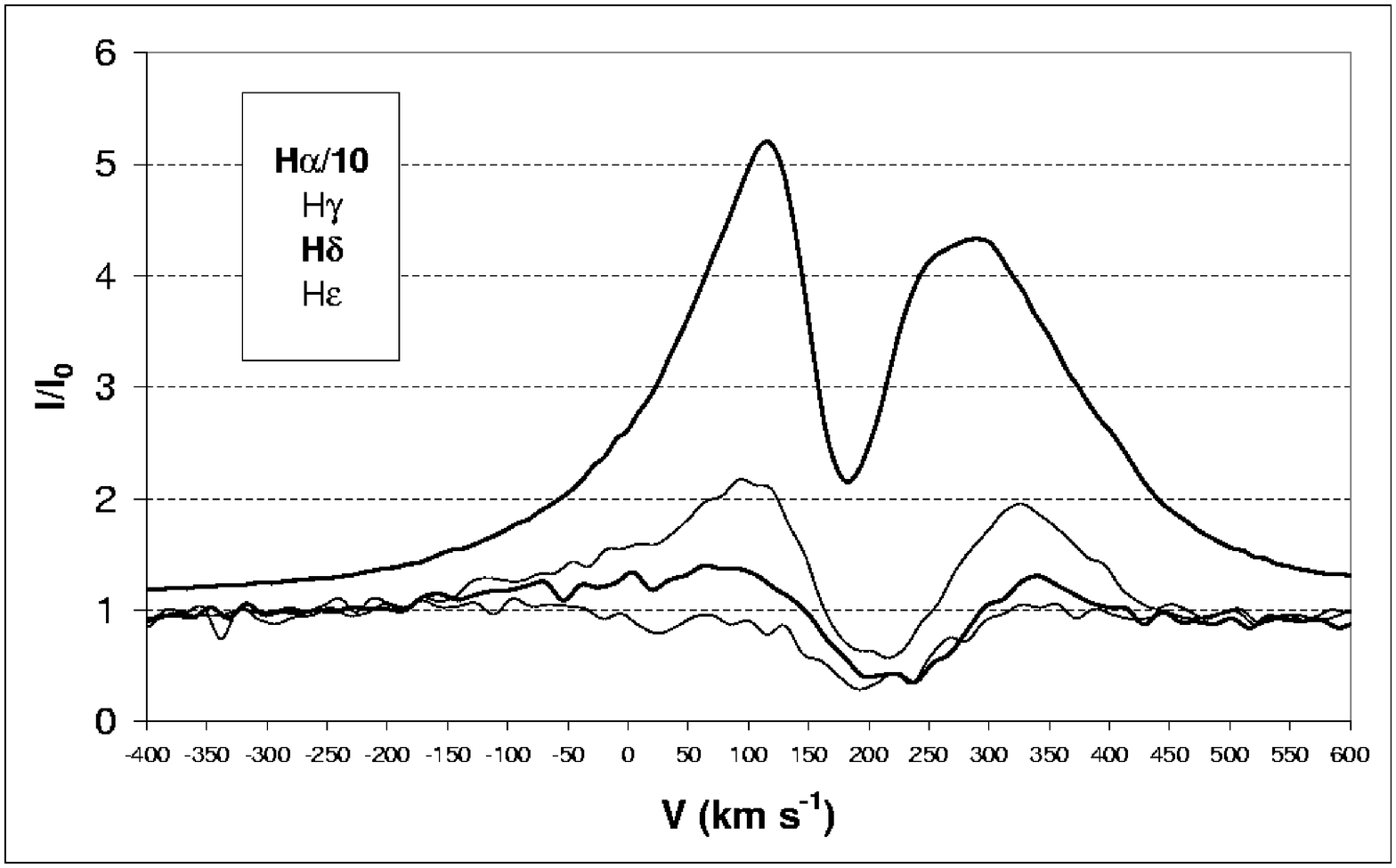}
\caption{Radial velocities of H$\alpha$, H$\gamma$, H$\delta$ and H$\epsilon$ for
BAL224.}
\label{vitesses}
\end{figure}

Two spectra of BAL224 ($\alpha$(2000), $\delta$(2000): 00h 56mn 06.45s, -72$^{o}$ 28' 27.70") were obtained at medium
resolution in setups LR02 (396 - 457 nm, R=6400) and LR06 (644 - 718 nm,
R=8600). 
They are dominated by the 2-peak emission components of Balmer lines
which are strongly asymmetric with V$>$$>$R. Due to the resolution used it was possible,
for the first time, to identify emission lines of [FeII], FeII, [CrI] as well as nebular
lines [SII]6717, 6731 (see Fig~\ref{specL26}). The mean radial velocity of these lines (RV)
is 154 \kms. The FWHMs of metallic emission lines are about 100
\kms~ and correspond to the instrumental broadening. The low S/N ratio in the continuum
(S/N$\simeq$20) did not allow to measure the radial velocity of HeI lines present in the
spectra of BAL224. The RVs mean values of the shell component of H$\alpha$, H$\gamma$
and H$\delta$ (see Figs~\ref{specL26}, \ref{vitesses} and Table~\ref{vitrad}) is 187 \kms. 

\begin{table*}[tbph]
\caption{Observational indications such as radial velocities or intensities of lines in the spectra of BAL224.
The values between brackets come from Hummel et al. (1999).}
\centering
\begin{tabular}{@{\ }c@{\ \ \ }c@{\ \ \ }c@{\ \ \ }c@{\ \ \ }c@{\ \ \ }c@{\ \ \ }c@{\ \ \ }c@{\ \ \ }c@{\ \ \ }c@{\ }}
\hline
\hline	
	& H$\alpha$ & H$\gamma$ & H$\delta$ \\
\hline	
RV$_{V}$ ($\pm$20) \kms & 104 [140 $\pm$50] & 86 & 62 \\
RV$_{shell}$ ($\pm$20) \kms & 171 & 198 & 204 \\
RV$_{R}$ ($\pm$20) \kms & 276 [301 $\pm$50] & 317 & 327 \\
FWHM ($\pm$20) \kms & 320 [443 $\pm$50] & 410 & 600 \\
I$_{V}$ & 41.8 & 2.2 & 1.4 \\
I$_{R}$ & 33.4 & 1.9 & 1.3 \\
Mean I & 37.6 & 2.1 & 1.4 \\
EW ($\pm$20) \AA & 360 [202 $\pm$20] & & \\
\hline
Ratios & H$\gamma$/H$\alpha$=0.055 & H$\delta$/H$\alpha$=0.036 & H$\delta$/H$\gamma$=0.66\\
\hline
\end{tabular}
\label{vitrad}
\end{table*}
\section{Photometric Variability}
\begin{figure*}[!ht]
\centering
\includegraphics[width=5cm, height=10cm,angle=-90]{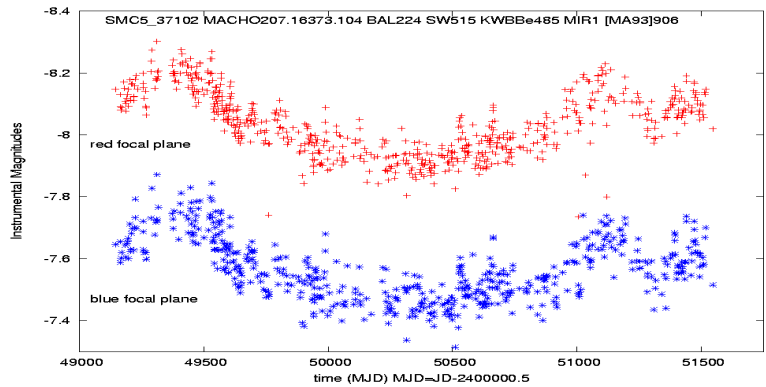}
\caption{Light-curve of BAL224 from MACHO database.}
\label{photvar}
\end{figure*}

According to Balona (1992), this star displayed fading of 0.2 mag and periods
close to 1 day but none of these periods could fit satisfactorily the data. 
Thanks to the MACHO and OGLE databases, 2 strong bursts  (Fig.~\ref{photvar})
could be observed with an amplitude of 0.4 mag on a time scale of about 3100
days. Between these 2 strong bursts, smaller ones which do not seen to be
periodic could also be observed. We searched for short-term variability and like
in Balona (1992) we find periods close to 1 day which do not  give a
satisfactory fit of the data. But irregular short- and long-term variabilities
may also be explained by the presence of a multiple object. 

\section{On the nature of BAL224}
From VLT-FORS1 low resolution spectroscopic observations Hummel et al. (1999)
suggest that the absence of emission in HeI lines and the strong Balmer decrement
can indicate that this star has a shell with a gas cooler than 5000K. Kucinskas et
al (2000) thanks to their photometric study found a strong mid-IR excess
compatible with a dust shell with a very low temperature: 360K. This infrared
excess is compatible with B[e] and Herbig stars but the temperature determined is not
compatible with B[e] stars. We confirm a strong Balmer decrement. No emission components
can be observed on HeI lines and some lines of neutral elements such as [CrI] are
present so we can conclude that a cool dust shell is present (Table~\ref{vitrad},
Figs~\ref{specL26}, \ref{vitesses}). The presence of FeII and [FeII] and their
FWHM lower than 100 \kms~ are common points between B[e] and Herbig B[e] stars.
But, we find an EW(H$\alpha$) smaller than 1000 \AA, which does not correspond to a
B[e]. The H$\alpha$ spectrum seen in Hummel et al (2000) and in this study clearly
shows a strong asymmetric double peak which may be explained by an accretion disk
(Fig.~\ref{vitesses}). This type of disk is a main characteristic of Herbig objects. Moreover,
the short- and long-term irregular variabilities are characteristic of Herbig
objects which may be explained by an aggregate of stars. Properties of B[e]
supergiants, HAeBe and isolated HAeBe (or HB[e]) are compared with properties of
BAL224 in Table~\ref{nature}. From this comparison, \textbf{we propose BAL224 as an isolated Herbig
B[e] object.}

\begin{table*}[tbph]
\scriptsize{
\caption{Comparisons between properties of: a B[e] supergiant (Sg), a Herbig Be (HAeBe), an isolated Herbig Be or HB[e] and BAL224.}
\centering
\begin{tabular}{@{\ }c@{\ \ \ }c@{\ \ \ }c@{\ \ \ }c@{\ \ \ }c@{\ \ \ }c@{\ \ \ }c@{\ \ \ }c@{\ \ \ }c@{\ \ \ }c@{\ }}
\hline
\hline	
Properties & B[e] Sg & HAeBe & HB[e] & BAL224\\
\hline	
FeII and [FeII] lines in emission & Yes & & Yes & Yes (this study)\\
FWHM FeII, [FeII]$<$100\kms & Yes & & & Yes (this study)\\
EW H$\alpha$ $>$1000\AA & Yes & & & No (this study + Hummel et al. 1999)\\
Near or far IR excess & Yes & Yes & Yes & Yes (Sebo \& Wood 1994)\\
IR excess, T$_{envelope}$$>$1000K & Yes & & & No (Kucinskas et al. 2000)\\
Excretion disk & Yes & & & No (this study + Hummel et al. 1999)\\
In obscure region & & Yes & & No (Balona 1992)\\
A-type or earlier & &  & & \\
+ emission lines & & Yes & & Yes (this study + Hummel et al. 1999)\\
Star illuminates nebulosity & &  &  &  \\
in immediate vicinity & & Yes & Yes & ? \\
Accretion disk & & Yes & Yes & Yes (this study + Hummel et al. 1999)\\
Irregular variations & & Yes & Yes & Yes (this study + Balona 1992)\\
Isolated object & & & Yes & Yes (Balona 1992)\\
Center of small aggregates & & & & \\
of low-mass stars & & & Yes & ? \\
\hline
\end{tabular}
\label{nature}
}
\end{table*}
%




\end{document}